\documentclass[preprint2]{aastex6}

\usepackage{natbib}
\usepackage{amsmath}
\usepackage{braket}
\begin{document}
\def\vec#1{\mbox{\boldmath $#1$}}

\title{A self-consistent model of the coronal heating and solar wind acceleration including compressible and incompressible heating processes}

\author{Munehito Shoda\altaffilmark{1}}
\author{Takaaki Yokoyama\altaffilmark{1}}
\altaffiltext{1}{Department of Earth and Planetary Science, The University of Tokyo, Hongo, Bunkyo-ku, Tokyo, 113-0033, Japan}
\author{Takeru K. Suzuki\altaffilmark{2}}
\altaffiltext{2}{School of Arts \& Sciences, The University of Tokyo, 3-8-1, Komaba, Meguro, Tokyo, 153-8902, Japan}

\begin{abstract}

	We propose a novel one-dimensional model that includes both shock and turbulence heating and qualify how these processes contribute to heating the corona and driving the solar wind.
	Compressible MHD simulations allow us to automatically consider shock formation and dissipation,
	while turbulent dissipation is modeled via a one-point closure based on Alfv\'en wave turbulence.
	Numerical simulations were conducted with different photospheric perpendicular correlation lengths $\lambda_0$, which is a critical parameter of Alfv\'en wave turbulence, 
	and different root-mean-square photospheric transverse-wave amplitudes $\delta v_0$.
	For the various $\lambda_0$, we obtain a low-temperature chromosphere, high-temperature corona, and supersonic solar wind.
	Our analysis shows that turbulence heating is always dominant when $\lambda_0 \lesssim 1{\rm \ Mm}$.
	This result does not mean that we can ignore the compressibility because the analysis indicates
	that the compressible waves and their associated density fluctuations enhance the Alfv\'en wave reflection and therefore the turbulence heating.
	The density fluctuation and the cross helicity are strongly affected by $\lambda_0$, while the coronal temperature and mass loss rate depend weakly on $\lambda_0$.
	
\end{abstract}


\section{Introduction} \label{sec:introduction}

	The mechanism sustaining the high-temperature corona \citep{Edlen43} and solar wind \citep{Parke58,Velli94} is still under investigation.
	Currently, it is widely accepted that the original energy source lies in the photpspheric convective motion and its interaction with magnetic fields \citep{Alfve47,Oster61,Stein98,Balle98,Fujim09,Kato011,Kato016}.
	Since the existence of Alfv\'en waves is indicated by in-situ observations \citep{Belch71a,Bale005} and remote sensing \citep{DePon07a,Tomcz07,Okamo11,McInt11,Thurg14},
	Alfv\'en-wave-driven models of the coronal heating \citep{Barne69,Cranm99} and solar wind acceleration \citep{Belch71b,Jacqu77} have been studied.
	Such models can explain the heating of both the coronal holes \citep{Hollw86,Suzuk05,Verdi10,Lione14,Matsu14,Balle16} and coronal loops \citep{Moriy04,Antol08,Balle11,Verdi12a}.
	Some models can explain the formation of both fast and slow solar winds \citep{Suzuk06a,Cranm07} by varying the expansion factor \citep{Wang090,Arge000}.
	
	To better model the corona and solar wind, the Alfv\'en wave propagation and dissipation needs to be solved precisely because the obtained structure is dependent on the heating distribution, that is, heating below and above the sonic point leads to slow and fast wind, respectively \citep{Hanst12}.
	To take into account heating mechanisms such as parametric decay \citep{DelZa01,DelZa15,Shi0017}, Alfv\'en wave turbulence \citep{Irosh64,Kraic65,Velli89,Goldr95,Matth99} and phase mixing \citep{Heyva83},
	we need to solve the three-dimensional compressible MHD equations with a sufficiently high resolution.
	However, due to its numerical difficulty and physical complexity, no models have ever included all the heating processes.
	To simplify the physics and reduce the computational cost, reduced MHD models have frequently been used \citep{Matth99,Dmitr02,Ought06,Cranm07,Verdi09,Chand09a,Verdi10,Perez13,Lione14,Balle16}.
	In such models, Alfv\'en wave turbulence driven by partial wave reflection \citep{Ferra58,Heine80,An00090,Velli93,Hollw07} is the only heating process.
	Some reflection-driven Alfv\'en wave turbulence models self-consistently succeed in explaining the coronal heating and solar wind accelerations \citep{Cranm07,Verdi10}.
	Recent 3D models, however, indicate that, under a smooth background field, turbulence cannot supply enough energy to sustain the corona \citep{Perez13,Balle17}.
	This indicates that we need to take into account the compressibility of plasma.
	In fact, compressible MHD models excluding Alfv\'en wave turbulence also succeed in generating the corona and solar wind \citep{Suzuk04,Suzuk05} based on shock heating.
	To summarize these recent studies, we need to solve for both shock (compressibility) and turbulence to achieve realisitc models.
	In this study, to solve both shock and turbulence heating without expensive numerical cost, we constructed a new 1D model of the coronal heating and solar wind acceleration.
	We solve the 1D compressible MHD equations to include shock formation and dissipation, and a one-point closure model is incorporated to include turbulence heating.
	The one-dimensionality and the resulting low numerical cost enable us to conduct a parameter survey to achieve a better understanding of the relevant physics.
	In this study, we investigate which type of heating is dominant in the corona and solar wind.
	
	The remainder of this paper is organized as follows.
	In Section \ref{sec:method}, we discuss how to incorporate turbulent dissipation into one-dimensional MHD equations. 
	The basic equations and numerical solver are also described in this section.
	The results are shown in Section \ref{sec:result}, and we summarize the paper with a discussion in Section \ref{sec:summary_discussion}.
	
\section{Method} \label{sec:method}

\subsection{Phenomenological terms of Alfv\'en wave turbulence}\label{sec:method:dissipation}

        We consider a one-dimensional system with its coordinate curved along the background flux tube.
        We denote the coordinate along the tube by $r$; therefore, $\partial/\partial r \ne 0$.
        In this case, following \citet{Suzuk05,Suzuk06a}, the basic equations for the transverse velocity $\vec{v}_\perp$ and magnetic field $\vec{B}_\perp$ are given as
        \begin{align}
          &\rho \frac{d}{dt} \left( \sqrt{A} \vec{v}_\perp \right) = \frac{B_r}{4 \pi} \frac{\partial}{\partial r} \left( \sqrt{A} \vec{B}_\perp \right), \\
          &\frac{\partial}{\partial t} \vec{B}_\perp = \frac{1}{\sqrt{A}} \frac{\partial}{\partial r} \left[ \sqrt{A} \left( \vec{v}_\perp B_r - v_r \vec{B}_\perp \right) \right],
        \end{align}
        where, in this subsection, we use $A$ instead of $r^2 f$ to denote the cross section of the tube.
        $d/dt$ represents the Lagrangian derivative: $d/dt = \partial/\partial t + v_r \partial/\partial r$.
        Using the mass conservation and solenoidal condition,
        \begin{align}
          \frac{d}{dt} \rho + \frac{\rho}{A} \frac{\partial}{\partial r} \left( A v_r \right) = 0, \ \ \ \ \frac{\partial}{\partial r} \left( A B_r \right) = 0,
        \end{align}
        and taking into account the turbulent diffusivity, the equations for $\vec{v}_\perp$ and $\vec{B}_\perp$ are written as
        \begin{align}
          &\frac{\partial}{\partial t} \left( \rho \vec{v}_\perp A^{3/2} \right)
          + \left[ \left( \rho v_r \vec{v}_\perp - \frac{1}{4 \pi} B_r \vec{B}_\perp \right) A^{3/2} \right] \nonumber \\
          & = - \hat{\vec{\eta}}_1 \cdot \rho \vec{v}_\perp A^{3/2} - \hat{\vec{\eta}}_2 \cdot \sqrt{\frac{\rho}{4 \pi}} \vec{B}_\perp A^{3/2} , \label{eq:vperp_mod} \\
          &\frac{\partial}{\partial t} \left( \vec{B}_\perp A^{1/2} \right)
          + \frac{\partial}{\partial r} \left[ \left( v_r \vec{B}_\perp - \vec{v}_\perp B_r \right) A^{1/2} \right] \nonumber \\
          & = -  \hat{\vec{\eta}}_1 \cdot \vec{B}_\perp A^{1/2} - \hat{\vec{\eta}}_2 \cdot \sqrt{4 \pi \rho} \vec{v}_\perp A^{1/2}, \label{eq:bperp_mod}
        \end{align}
        where $\hat{\vec{\eta}}_1$ and $\hat{\vec{\eta}}_2$ are diagonal matrices whose components are given as
        \begin{align}
          &\hat{\vec{\eta}}_1=
          \left(
          \begin{array}{cc}
            \dfrac{c_d}{4 \lambda} \left( |\zeta^+_x| + |\zeta^-_x| \right) & 0 \\
            0 & \dfrac{c_d}{4 \lambda} \left( |\zeta^+_y| + |\zeta^-_y| \right) \\
          \end{array}
          \right), \\
          &\hat{\vec{\eta}}_2=
          \left(
          \begin{array}{cc}
            \dfrac{c_d}{4 \lambda} \left( |\zeta^+_x| - |\zeta^-_x| \right) & 0 \\
            0 & \dfrac{c_d}{4 \lambda} \left( |\zeta^+_y| - |\zeta^-_y| \right) \\
          \end{array}
          \right),
        \end{align}
        where $\lambda$ is the correlation length perpendicular to the mean field, $x$ and $y$ denote the two transverse components, and $\zeta$ represents the Els\"asser variables \citep{Elsas50} given by
        \begin{align}
          \vec{\zeta^{\pm}} = \vec{v}_\perp \mp \frac{\vec{B}_\perp}{\sqrt{4 \pi \rho}}.
        \end{align}
        Physically, $\hat{\vec{\eta}}_1$ represents a simple dissipation by turbulence, while $\hat{\vec{\eta}}_2$ comes from a relaxation process called dynamic alignment \citep{Dobro80,Strib91}.
	In fact, $\hat{\vec{\eta}}_2$ arises only for imbalanced turbulence \citep{Berez08,Chand08}.

        Here, we show that, as long as the density and radial velocity are time-independent as is often assumed,
        Eqs. (\ref{eq:vperp_mod}) and (\ref{eq:bperp_mod}) are consistent with the common Alfv\'en wave turbulence term.
        Introducing $\vec{b} = \vec{B}/\sqrt{4 \pi \rho}$, we can rewrite eq.s (\ref{eq:vperp_mod}) and (\ref{eq:bperp_mod}) as
        \begin{align}
          &\frac{\partial}{\partial t} \vec{v}_\perp + v_r \frac{\partial}{\partial r} \vec{v}_\perp - b_r \frac{\partial}{\partial r} \vec{b}_\perp \nonumber \\
          + &v_r \vec{v}_\perp \frac{\partial}{\partial r} \ln \left(  A^{1/2} \right) - b_r \vec{b}_\perp \frac{\partial}{\partial r} \ln \left( \rho^{1/2} A^{1/2} \right) \nonumber \\
          =& -\hat{\vec{\eta}}_1 \cdot \vec{v}_\perp - \hat{\vec{\eta}}_2 \cdot \vec{b}_\perp, \\
          &\frac{\partial}{\partial t} \vec{b}_\perp + v_r \frac{\partial}{\partial r} \vec{b}_\perp - b_r \frac{\partial}{\partial r} \vec{v}_\perp \nonumber \\
          + &b_r \vec{v}_\perp \frac{\partial}{\partial r} \ln \left(  A^{1/2} \right) - v_r \vec{b}_\perp \frac{\partial}{\partial r} \ln \left( \rho^{1/2} A^{1/2} \right) \nonumber \\
          =& -\hat{\vec{\eta}}_1 \cdot \vec{b}_\perp - \hat{\vec{\eta}}_2 \cdot \vec{v}_\perp.
        \end{align}
        These equations are written in Els\"asser variables as 
        \begin{align}
          &\frac{\partial}{\partial t} \vec{\zeta}^{\pm} + \left( v_r \pm b_r \right) \frac{\partial}{\partial r} \vec{\zeta}^{\pm}, \nonumber \\
          - & \vec{\zeta}^{\pm} \left(v_r \mp b_r \right) \frac{\partial}{\partial r} \ln \left( \rho^{1/4} \right)
          + \vec{\zeta}^{\mp} \left( v_r \mp b_r \right) \frac{\partial}{\partial r} \ln \left( \rho^{1/4} A^{1/2} \right) \nonumber \\
          &= - \hat{\vec{\eta}}^{\mp} \cdot \vec{\zeta}^{\pm}, \label{eq:ho80_mod}
        \end{align}
        where
        \begin{align}
          \hat{\vec{\eta}}^{\mp} = \hat{\vec{\eta}}_1 \mp \hat{\vec{\eta}}_2 =
          \left(
          \begin{array}{cc}
            \dfrac{c_d}{2 \lambda} |\zeta^\mp_x| & 0 \\
            0 & \dfrac{c_d}{2 \lambda}  |\zeta^\mp_y| \\
          \end{array}
          \right).
        \end{align}
        Note that $v_r$ and $b_r$ represent the background flow speed and the Alfv\'en velocity, respectively.
        The left hand side of Eq. (\ref{eq:ho80_mod}) is identical to Eqs. (4a) and (4b) in \citet{Heine80}.
        The right hand side is equivalent to the Alfv\'en wave turbulence term approximated as \citep{Dobro80,Hossa95,Matth99}
        \begin{align}
		- \left( \vec{\zeta}^{\mp} \cdot \vec{\nabla} \right) \vec{\zeta}^{\pm} \sim - c_d \frac{Z^{\mp}}{2\lambda} \vec{\zeta}^{\pm},
		\label{eq:dissipation}
	\end{align}
        where $Z^{\mp}$ is the rms amplitude of $\zeta^{\mp}$, which is, following \citet{Chand09a} and \citet{Lione14}, approximated by $|\zeta^{\mp}|$ in this study.
        $\tau_{\rm turb}^{\pm} \sim \lambda/\zeta^{\mp}$ represents the timescale of the turbulence.
        The turbulence term $- \hat{\vec{\eta}}^{\mp} \cdot \vec{\zeta}^{\pm}$
        arises only when there exist counter-propagating Alfv\'en waves \citep{Irosh64,Kraic65}.
        The choice of $c_d$ is not trivial.
        Because the reflection timescale $\tau_{\rm ref}$ must be smaller than the nonlinear timescale $\tau_{\rm nl}$ to sustain the turbulence \citep{Dmitr03}, 
	\citet{Ought06} and \citet{Cranm07} evaluated $c_d$ as a function of the nonlinear timescale $\tau_{\rm nl}=\tau_{\rm turb}^{\pm}$ and the reflection timescale $\tau_{\rm ref}=\nabla \cdot \vec{V}_A$, 
	which gives $c_d = 0$ for $\tau_{\rm nl} / \tau_{\rm ref} \to \infty$ and $c_d = 1$ for $\tau_{\rm nl} / \tau_{\rm ref} \to 0$.
	Meanwhile, \citet{Dmitr02}, \citet{Verdi07}, \citet{Chand09a}, \citet{Verdi10}, and \citet{Lione14} simply assume $c_d =1$.
 	Both models give similar results for the heating and acceleration, because $c_d$ is approximately unity for the main heating region in both cases.

        However, it has recently been pointed out by \citet{Balle11} and \citet{Balle16} that
        this formulation overestimates the turbulent dissipation than the value obtained from 3D calculations.
	\citet{Perez13} report that the heating rate calculated via 3D RMHD simulations is smaller compared to the required value.
	\citet{Balle17} conclude that $c_d=0.1$ gives a better approximation for the heating rate.
	Therefore, we use $c_d=0.1$ in this study.
	Note that the actual value or function of $c_d$ should have a complex form that depends on the structure of magnetic field and power spectrum, and thus $c_d$ is also a free parameter.
	However, because increasing $c_d$ is mathematically equivalent with decreasing $\lambda$, we fixed $c_d$ with a reasonable value by \citet{Balle17} and changed $\lambda$ in this study.

\subsection{Basic equations}

        In addition to Eqs. (\ref{eq:vperp_mod}) and (\ref{eq:bperp_mod}), we solve the mass conservation equation, the equation of radial motion, and the energy equation.
        The basic equations in conservation form are
	
	\begin{align}
		&\frac{\partial}{\partial t} \left( \rho r^2 f \right) + \frac{\partial }{\partial r} \left( \rho v_r r^2 f \right) =0,  \label{eq:mass} \\ 
    		&\frac{\partial}{\partial t} \left( \rho v_r r^2 f \right) + \frac{\partial }{\partial r} \left[ \left( \rho {v_r}^2 + p + \frac{{\vec{B}_{\perp}}^2}{8\pi} \right) r^2 f \right] \nonumber \\
    		&= \left( p + \frac{\rho {\vec{v}_{\perp}}^2}{2} \right) \frac{d}{dr} \left( r^2 f \right) - \rho g r^2 f, \label{eq:eomradial} \\
    		&\frac{\partial}{\partial t} \left( \rho \vec{v}_{\perp} r^3 f^{3/2} \right) + \frac{\partial }{\partial r} \left[ \left( \rho v_r \vec{v}_{\perp} - \frac{B_r \vec{B}_{\perp}}{4 \pi} \right) r^3 f^{3/2} \right] \nonumber \\
   		&= -\hat{\vec{\eta}}_1 \cdot \rho \vec{v}_{\perp} r^3 f^{3/2} - \hat{\vec{\eta}}_2 \cdot \sqrt{\frac{\rho}{4 \pi}} \vec{B}_{\perp} r^3 f^{3/2}, \label{eq:vt} \\
    		&\frac{\partial}{\partial t} \left(\vec{B}_{\perp}  r \sqrt{f}  \right) + \frac{\partial }{\partial r} \left[ \left( \vec{B}_{\perp} v_r - B_r \vec{v}_{\perp} \right) r \sqrt{f} \right] \nonumber \\
		&= -\hat{\vec{\eta}}_1 \cdot \vec{B}_{\perp} r \sqrt{f} - \hat{\vec{\eta}}_2 \cdot \sqrt{4 \pi \rho} \vec{v}_{\perp} r \sqrt{f}, \label{eq:bt} \\
   	 	&\frac{d}{dr} \left(B_r  r^2 f  \right) = 0, \label{eq:divb} \\
  		&\frac{\partial}{\partial t} \left[ \left( e + \frac{1}{2} \rho \vec{v}^2 + \frac{\vec{B}^2}{8 \pi} \right) r^2 f \right] \nonumber \\
  		&+ \frac{\partial }{\partial r} \left[\left( e + p + \frac{1}{2} \rho \vec{v}^2  + \frac{{\vec{B}_{\perp}}^2}{4 \pi} \right ) v_r r^2 f - B_r \frac{\vec{B}_{\perp} \cdot \vec{v}_{\perp}}{4 \pi} r^2 f \right] \nonumber \\
    		&= r^2 f \left( - \rho g v_r + Q_{\rm rad} + Q_{\rm cond}  \right), \label{eq:energy} \\
		& e = \frac{p}{\gamma-1}, \ \ \ \ p = \frac{\rho k_B T}{\mu m_H}, \label{eq:eos}
	\end{align}
	where $r$ is the heliocentric distance and $f$ is the expansion factor of the flux tube \citep{Kopp076,Wang090}.
	See \citet{Shoda18} for derivation.
        Note that Eqs. (\ref{eq:vt}) and (\ref{eq:bt}) are equivalent to Eqs. (\ref{eq:vperp_mod}) and (\ref{eq:bperp_mod}).
	$Q_{\rm rad}$ and $Q_{\rm cond}$ denote the radiative cooling and thermal conduction, respectively.
	The gravitational acceleration is $g = 2.74 \times 10^4 \left( r/R_{\rm Sun} \right)^{-2} {\rm \ cm \ s^{-2}} $.
	We assume that the solar atmosphere is composed of only hydrogen.
	$\mu$ is given as a function of density so that $\mu=1$ in the photosphere and low chromosphere and $\mu=0.5$ in the high chromosphere and corona.
	We ignored the dependence of $\mu$ on temperature for simplicity. 
	Because we are interested in the coronal physics, and $\mu = 0.5$ (fully ionized) for coronal mass density, our assumption should not affect the numerical results.
			
	Following \citet{Suzuk05}, we assume a two-step super-radial expansion of the flux tube such that
	\begin{align}
		f = \frac{f_{\rm max,1} \exp \left( \frac{r - R_1}{\sigma_1} \right) + f_1}{\exp \left( \frac{r - R_1}{\sigma_1} \right)+1} 
		\cdot \frac{f_{\rm max,2} \exp \left( \frac{r - R_2}{\sigma_2} \right) + f_2}{\exp \left( \frac{r - R_2}{\sigma_2} \right)+1},
	\end{align}
	where
	\begin{align}
		f_1 &= 1 - \left( f_{\rm max,1} -1 \right) \exp \left( \frac{R_{\rm Sun} - R_1}{\sigma_1} \right), \nonumber \\
		f_2 &= 1 - \left( f_{\rm max,2} -1 \right) \exp \left( \frac{R_{\rm Sun} - R_2}{\sigma_2} \right). \nonumber
	\end{align}
	The first and second terms represent the chromospheric and coronal expansions, respectively.
	We assume that the first expansion is near $r-R_{\rm Sun} = 1 {\rm \ Mm}$ with a length scale of $\sigma_1 = 0.25 {\rm \ Mm}$ while the second is near $r-R_{\rm Sun} = 200 {\rm \ Mm}$ with a length scale of $\sigma_2 = 350 {\rm \ Mm}$.
	The expansion factors  are $f_{\rm max,1}=120$ and $f_{\rm max,2} = 4$.
	The background magnetic field is calculated from $f$ as
	\begin{align}
		B_r = B_{r,0} \left( \frac{r}{R_{\rm Sun}} \right)^{-2} f^{-1}.
	\end{align}
	The correlation length is assumed to increase with the expansion of the flux tube \citep{Dmitr02}.
	In addition, to restrict the Alfv\'en wave turbulence to the corona \citep{Cranm07,Verdi10}, we formulate $\lambda$ such that
	\begin{align}
		\lambda  = \lambda_0 \sqrt{\frac{B_{r,0}}{B_r}} \max \left(1,\frac{\rho}{\rho_{\rm cor,1}} \right)  \label{eq:correlation},
	\end{align}
	where $\rho_{\rm cor,1} = 10^{-16} {\rm \ g \ cm^{-3}}$.
	This results in a turbulent dissipation slow enough to be negligible in the chromosphere.
	Note that some models indicate that the Alfv\'en wave turbulence is active also in the chromosphere \citep{Balle11,Verdi12a}.
	The purpose of this assumption is to compare our results to those of previous studies; there is no physical justification to exclude chromospheric Alfv\'en wave turbulence.
	
	Radiative cooling is a combination of optically thick $L_{\rm thick}$ and thin $L_{\rm thin}$ contributions and is given by
	\begin{align}
		Q_{\rm rad} = \xi L_{\rm thick} + (1-\xi) L_{\rm thin},
	\end{align}
	where 
	\begin{align}
		\xi = \max \left[0,\left(1-\frac{\rho_{\rm cor,2}}{\rho} \right) \right].
	\end{align}
	Here, $\rho_{\rm cor,2} = 5 \times 10^{-17} {\rm \ g \ cm^{-3}}$.
	Following \citet{Gudik05}, instead of solving the radiative transfer, we approximate the optically thick cooling by Newtonian cooling as
	\begin{align}
		L_{\rm thick} = -\frac{1}{\tau_{\rm thick}} \left( e - e_0 \right),
	\end{align}
	where 
	\begin{align}
		\tau_{\rm thick} = 1 \times \left( \frac{\rho}{\rho_0} \right)^{-1} {\rm \ s}, \ \ \ \ e_0 = \frac{1}{\gamma-1} \frac{\rho k_B T_{\rm ref}}{\mu m_H}.
	\end{align}
	Here, $\rho_0$ is the photospheric mass density and $\tau_{\rm thick}$ denotes the timescale of the cooling.
        $T_{\rm ref}$ represents a reference temperature that sets $T_{\rm ref}=6000 {\rm \ K}$ at $r = R_{\rm Sun}$ and monotonically decreases with $r$.
	As for $L_{\rm thin}$, we use the following formula:
	\begin{align}
  		L_{\rm thin} = - n_i n_e \Lambda(T), \ \ \ \ n_i = n_e = \frac{\rho}{m_H} \left(\frac{1}{\mu} -1 \right),
	\end{align}
	where we use an approximated loss function for $\Lambda(T)$ \citep{Suthe93,Matsu14}.

	In the chromosphere and low corona where the plasma is collisional, Spitzer-H\"arm conduction \citep{Spitz53} is applicable.
	However, in the solar wind, Spitzer-H\"arm conduction is inadequate due to the long mean free path, and thus free-streaming type conducton \citep{Hollw74,Hollw76} should be used instead.
	Because the free-streaming conductive flux is smaller than the Spitzer-H\"arm conductive flux, to mimic the transition to free-streaming conduction, 
	we formulate the conductive heating $Q_{\rm cond}$ and conductive flux $q_{\rm cond}$ such that
	\begin{align}
		Q_{\rm cond} = - \frac{1}{r^2 f} \frac{\partial}{\partial r} \left( \alpha q_{\rm cond} r^2 f \right)
	\end{align}
	where
	\begin{align}
		q_{\rm cond} = - \kappa T^{5/2} \frac{\partial T}{\partial r}, \ \ \ \ \alpha = \max \left(1,\frac{\rho}{\rho_{\rm sw}} \right)
	\end{align}
	where $\rho_{\rm sw}=10^{-22} {\rm \ g \ cm^{-3}}$ and $\kappa=10^{-6}$ in CGS-Gaussian unit.
        $\alpha$ represents the flux quenching in the solar wind ($\rho < \rho_{\rm SW}$).
	
	\begin{figure}[!t]
	\vspace{3em}
	\begin{flushright}
	\includegraphics[width=80mm]{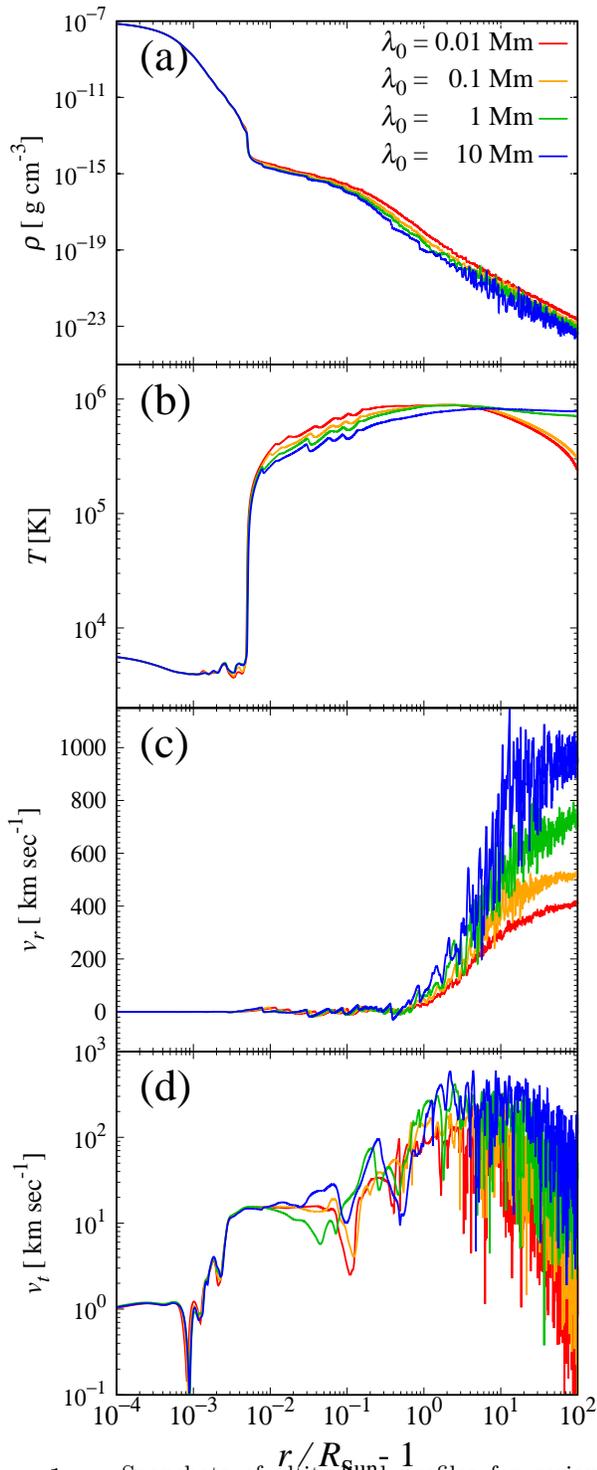}
	\end{flushright}
	\vspace{-3em}
	\caption{
		Snapshots of altitudinal profiles for various $\lambda_0$.
		The mass density, temperature, radial velocity and transverse velocity are shown from the top to bottom, 
		The red, orange, green and blue lines indicate 
		$\lambda_0= 0.01 {\rm \ Mm}$, $\lambda_0= 0.1 {\rm \ Mm}$, 
		$\lambda_0= 1 {\rm \ Mm}$, and $\lambda_0= 10 {\rm \ Mm}$, respectively.
										}
	\label{fig:result_compare}
	\end{figure}
		
\subsection{Boundary conditions and schemes}

        The basic equations are solved from the bottom (photosphere) up to $0.5 {\rm \ au}$ using 13000 non-uniform grid points.
        A static atmosphere with a temperature of $10000 {\rm \ K}$ is used as the initial condition \citep{Suzuk05}.
	The photospheric boundary conditions are as follows.
	The density, temperature, and radial magnetic field are fixed to $\rho_0 = 10^{-7} {\rm \ g \ cm^{-3}}$, $T_0 = 6000 {\rm \ K}$, and $B_{r,0} = 1000 {\rm \ G}$, respectively.
	The free boundary condition is applied to the radial velocity at the photosphere.
	The transverse velocity and magnetic field with pink-noise spectra are given at the photosphere so that upward waves are excited:
	\begin{align}
		\vec{v}_{\perp,0} &\propto \int_{2 \pi f_{\rm l}}^{2 \pi f_{\rm h}} d\omega \ \omega^{-1/2} \sin(\omega t + \phi(\omega) ), \\ 
		\vec{B}_{\perp,0}  &= - \sqrt{4 \pi \rho_0} \vec{v}_{\perp,0},
	\end{align}
	where $f_{\rm l}=10^{-3} {\rm \ Hz}$ and $f_{\rm h}=10^{-2} {\rm \ Hz}$.
        $\phi$ is a random function of $\omega$.
	The photospheric correlation length $\lambda_0$ and the root-mean-square value of $|\vec{v}_{\perp,0}|$ denoted by $\delta v_0$ are the free parameters in this study.
	As for the upper boundary condition, we apply the free boundary condition to every variable.
	Even though the transmitting boundary condition \citep{Thomp87,DelZa01,Suzuk06a} is likely better, we confirmed that the numerical result does not depend greatly on the boundary condition because a super-sonic and super-Alfv\'enic outflow is obtained as a final state.
	We solve the conservation law using an approximated Riemann solver \citep{Miyos05} with 2nd-order MUSCL reconstruction and 3rd-order TVD Runge-Kutta method \citep{Shu0088}.
	Super time stepping method \citep{Meyer12,Meyer14} is applied to solve the Spitzer-type thermal conduction in Eq. (\ref{eq:energy}), which drastically improves the calculation speed.

	\begin{figure}[!t]
	\vspace{3em}
	\begin{flushright}
	\includegraphics[width=80mm]{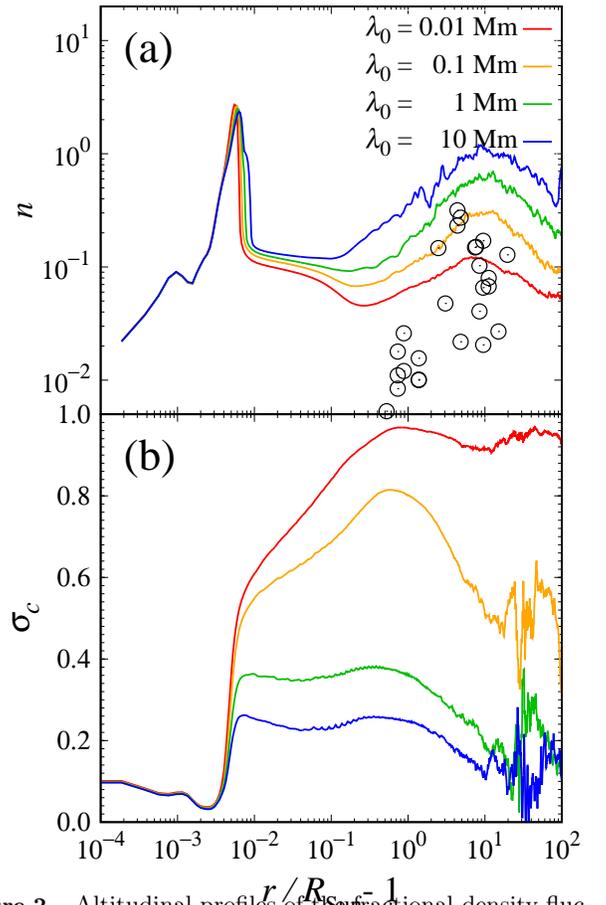}
	\end{flushright}
	\vspace{-3em}
	\caption{
		Altitudinal profiles of the fractional density fluctuation $n$ and the normalized cross helicity $\sigma_c$.
		Shown by the four lines are cases with $\lambda_0=0.01{\rm \ Mm}$ (red), 
		$\lambda_0=0.1{\rm \ Mm}$ (orange), $\lambda_0=1{\rm \ Mm}$ (green), and $\lambda_0=10{\rm \ Mm}$ (blue).
		Black circles indicate the radio-wave observations by \citet{Miyam14}.
				}
	\label{fig:ch_dro_compare}
	\end{figure}
	
\section{Result} \label{sec:result}

\subsection{Quasi-steady states for various $\lambda_0$ \label{subsec:qss}}

	Figure \ref{fig:result_compare} shows snapshots of the quasi-steady states for four cases with different $\lambda_0$ values. 
	The photospheric transverse velocity is fixed to $\delta v_0 = 0.5 {\rm \ km \ s^{-1}}$.
	The panels show profiles of (a) the mass density, (b) temperature, (c) radial velocity, and (d) transverse velocity $v_t = |\vec{v}_\perp|$.
	The four lines indicate the results with $\lambda_0=0.01 {\rm \ Mm}$ (red), 
	$\lambda_0=0.1 {\rm \ Mm}$ (orange), $\lambda_0=1 {\rm \ Mm}$ (green), and $\lambda_0=10 {\rm \ Mm}$ (blue).
	In each case, a chromosphere and corona are generated with a sharp transition region in-between them and a solar wind with velocity $\gtrsim 400 {\rm \ km\cdot sec^{-1}}$ is formed.
		
	Variations with respect to $\lambda_0$ will now be discussed.
	The solar wind has a different asymptotic velocity, which is larger for larger $\lambda_0$ (Figure \ref{fig:result_compare}c).
	The longer timescale of the turbulent dissipation makes waves propagate longer distances, giving rise to a faster solar wind because the heating in the supersonic region increases \citep{Lamer99,Hanst12}.
	The difference in the temperature profiles can also be interpreted via the location of the heating; the lower corona is hotter for smaller $\lambda_0$, while the wind is hotter for larger $\lambda_0$ (Figure \ref{fig:result_compare}b). 
        As for the wave amplitudes (Figure \ref{fig:result_compare}d), the four cases show a similar trend.
	We see $v_t = 10-30 {\rm \ km \ s^{-1}}$ near the transition region and low corona, which is consistent with the observational values \citep{DePon07a,McInt11,Thurg14}.
	An increasing trend of $v_t$ up to $r/R_{\rm Sun}-1 \sim 3$ is seen in every case and is consistent with recent observation of non-thermal line broadening \citep{Baner09,Hahn013}.

	The density fluctuation and associated radial velocity fluctuation strongly depends on $\lambda_0$.
	The large density fluctuation, which was observed also in previous studies \citep{Suzuk05,Suzuk06a}, 
	comes from a large amplitude MHD slow (acoustic) waves generated via the parametric decay instability \citep{Sagde69,Golds78,DelZa01}.
	This is clear from the relationship between the time-averaged fractional density fluctuation $n$ and the normalized cross helicity $\sigma_c$ defined as
	\begin{align}
		n &= \frac{1}{\bar{\rho}}\sqrt{\overline{(\rho-\overline{\rho})^2}}, \\
		\sigma_c &= \frac{\overline{E^{+}}-\overline{E^{-}}}{\overline{E^{+}}+\overline{E^{-}}}, \ \ \ \ E^{\pm} = \frac{1}{4} \rho \vec{\zeta^{\pm}}^2,
	\end{align}
	where $\overline{X}$ denotes the time average: $\overline{X} = 1/\tau \int^\tau_0 dt X$, where $\tau=1000{\rm \ min.}$
	In Figure \ref{fig:ch_dro_compare} we show the altitudinal profiles of $n$(a) and $\sigma_c$(b).
        It is clear, especially for large $\lambda_0$, that the increasing $n$ is associated with decreasing $\sigma_c$, which is consistent with the parametric decay instability.
        
        \begin{figure*}[!t]
	\begin{flushright}
	\includegraphics[width=175mm]{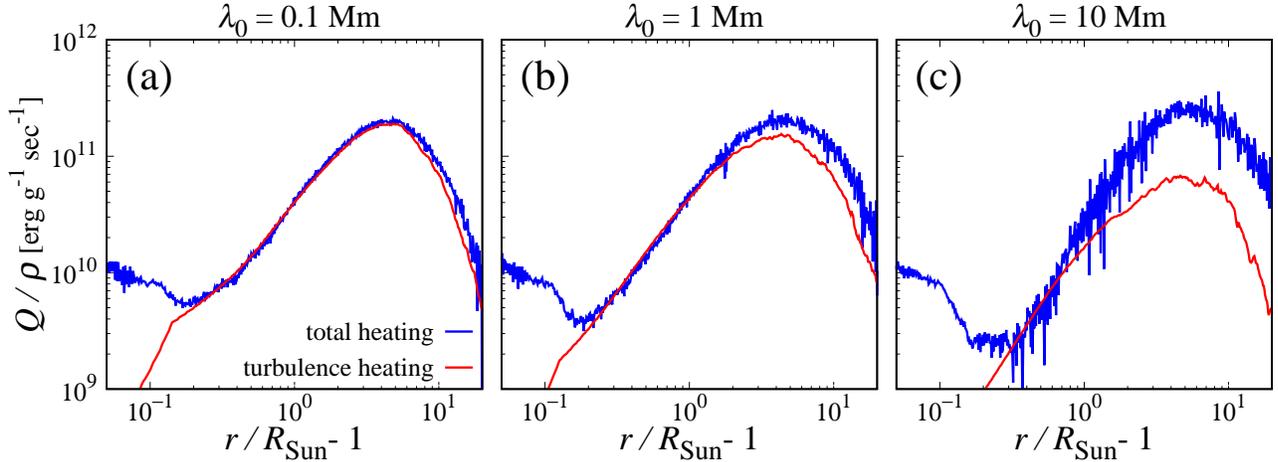} 
	\end{flushright}
	\vspace{1em}
	\caption{
		Altitudinal profiles of the time-averaged total heating ($Q_{\rm heat}/\rho$: blue line) and turbulence heating ($Q_{\rm turb}/\rho$: red line) for various $\lambda_0$.
		The left, middle and right panels indicate $\lambda_0=0.1{\rm \ Mm}$, $\lambda_0=1{\rm \ Mm}$, and $\lambda_0=10{\rm \ Mm}$, respectively.
				}
	\vspace{1em}
	\label{fig:heating_pergram_compare}
	\end{figure*}

        According to Figure \ref{fig:ch_dro_compare}a, $n$ has maxima near $r/R_{\rm Sun}-1=10^{-2}$ (the chromosphere and transition region) and $r/R_{\rm Sun}=10$ (the solar wind).
	The first one results from sound waves nonlinearly generated from Alfv\'en waves via the wave pressure gradient \citep{Hollw71,Hollw82a,Kudoh99,Matsu10}.
	The sound waves are amplified due to the stratification and steepen to form shock waves \citep{Carls92,Tian014}, leading to large density fluctuations.
	After a shock wave collides with the transition region,
        its energy is redistributed to the coronal shock wave, the upward motion of the transition region, and the chromospheric rarefaction wave \citep{Hollw82b}.
	This is the cause of the rapid decrease in $n$ above the transition region.
	The second peak results from the decay instability, which enhances $n$ and reduces $\sigma_c$ due to backscattering \citep{Malar00,DelZa01,Shoda16}.
	
	The black circles in Figure \ref{fig:ch_dro_compare}a indicate radio-wave observations by \citet{Miyam14} using {\it Akatsuki} \citep{Nakam11}.
        All four cases show an increasing trend for $n$ with $r$ in $0.1<r/R_{\rm Sun} -1 \lesssim10$, which is consistent with the observation.
	These observational data could underestimate the actual values because positive and negative density fluctuations, which cause radio scintillation, 
	are canceled out in the integration along the line of sight.
	Therefore, even though our theoretical results exceed most of the observational data, this discrepancy is not a contradiction.
	
	Even though the fluctuations in $\rho$ and $v_r$ has a large length scale typically of $0.1R_{\rm Sun}$, 
	they are not observable by remote sensing because the intensity of that region is extremely low.
	In-situ observations would be the only way to detect the fluctuations, which is expected from {\it Parker Solar Probe} \citep{Fox0016}.
	
	$n$ and $\sigma_c$ show different behaviors with different $\lambda_0$.
	First, the magnitude of $n$ near $r/R_{\rm Sun}=10$ is smaller for smaller $\lambda_0$.
	This is likely because the growth rate and saturation level of the decay instability decrease due to smaller wave powers or because the decay instability itself is suppressed by the presence of turbulence.
	Second, the magnitude of $\sigma_c$ in the corona and solar wind becomes smaller as $\lambda_0$ increases.
	This is a natural result of dynamic alignment \citep{Dobro80,Strib91}.
	Large $\sigma_c$ observed in the fast solar wind near $1{\rm \ au}$ \citep{Belch71a,Bavas00} likely results from small $\lambda_0$ ($\lambda_0 \lesssim 0.1 {\rm \ Mm}$) and the subsequent alignment.
	However, recent observations by CoMP reveal that a significant amount of reflected Alfv\'en waves exist in the low corona \citep{Morto15}.
	According to \citet{Morto15}, the energy ratio $E^- / E^+$ is $0.5-1.0$, which is, in terms of $\sigma_c$, $0-0.3$.
	Thus, $\lambda_0\gtrsim 1 {\rm \ Mm}$ is preferable to explain their observation.
	Explaining the cross helicity profile from the low corona up to the distant heliosphere $r\gtrsim 1{\rm \ au}$ is beyond the scope of this study and remains for future work.
			
\subsection{Heating mechanism
	\label{sec:heating_mechanism}}

	To measure the heating, we use the energy equation written in the following manner \citep{Cranm07}.
	\begin{align}
		\frac{\partial e}{\partial t} + v_r \frac{\partial e}{\partial r} + \frac{e+p}{r^2 f} \frac{\partial}{\partial r} \left( v_r r^2 f \right) = Q_{\rm rad} + Q_{\rm cond} + Q_{\rm heat},
	\end{align}
	where $Q_{\rm heat}$ is the heating by waves.
	This equation is derived from the basic equations.
	After a quasi-steady state is achieved, using time averaging, we obtain
	\begin{align}
		\overline{Q_{\rm heat}} = 
		\overline{v_r \frac{\partial e}{\partial r}  +\frac{e+p}{r^2 f} \frac{\partial}{\partial r} \left( v_r r^2 f \right) - Q_{\rm rad} - Q_{\rm cond}},
	\end{align}
        where we use $\overline{\partial e / \partial t}=0$.
	We can calculate the total wave heating in this manner.
	Meanwhile, the turbulent heating $Q_{\rm turb}$ is analytically derived as \citep{Verdi07,Cranm07}
	\begin{align}
		Q_{\rm turb} = \frac{1}{4} \rho \sum_{i=x,y} c_d \frac{|\zeta^{+}_i|{\zeta^{-}_i}^2+|\zeta^{-}_i|{\zeta^{+}_i}^2}{\lambda},
	\end{align}
	where $x$ and $y$ denote the components perpendicular to the background flux tube.
	In Figure \ref{fig:heating_pergram_compare}, we show the time-averaged heating per unit mass $\overline{Q_{\rm heat}}/\overline{\rho}$ (blue line) 
	and $\overline{Q_{\rm turb}}/\overline{\rho}$ (red line) for cases with the same wave amplitude ($\delta v_0 = 0.5 {\rm \ km \ s^{-1}}$) 
	and different correlation lengths ($\lambda_0=0.1 {\rm \ Mm}, 1{\rm \ Mm}, and 10{\rm \ Mm}$).
	In every case, the heating per unit mass reaches $\sim 10^{11} {\rm \ erg \ g^{-1} \ s^{-1}}$,
        showing that a sufficient amount of energy is supplied to sustain the coronal temperature.
	For $\lambda_0 \lesssim 0.1{\rm \ Mm}$, the turbulence heating is dominant everywhere; meanwhile for $\lambda_0 \gtrsim 1{\rm \ Mm}$, the shock heating is comparably or more important in the extended corona and solar wind.
	Thus our first conclusion is that the dominant heating process is turbulence when $\lambda_0 \lesssim 1{\rm \ Mm}$.
	In particular, when $\lambda_0 \lesssim 0.1 {\rm Mm}$, the shock heating is nearly negligible.
	What is interesting in this analysis is that, even though we apply small turbulence heating ($c_d = 0.1$),
        the sufficient amount of energy is supplied only by the turbulent heating when $\lambda_0= 1 {\rm \ Mm}$.
	In contrast, a reduced MHD calculation \citep{Perez13} gives a heating rate of at most $\sim 10^{10} {\rm \ erg \ g^{-1} \ s^{-1}}$ when $\lambda_0 \sim 1 {\rm \ Mm}$.
	The enhanced heating indicates that Alfv\'en wave reflection, which is the trigger of Alfv\'en wave turbulence, is activated due to the compressibility.
	This is possibly because of the presence of parametric decay or reflection driven by density fluctuations \citep{Suzuk06a,Balle16}.
	In this regard, the compressibility can never be ignorable not only due to shock heating but also due to reflection enhancement.
	
	\begin{figure*}[!t]
	\begin{flushright}
	\includegraphics[width=175mm]{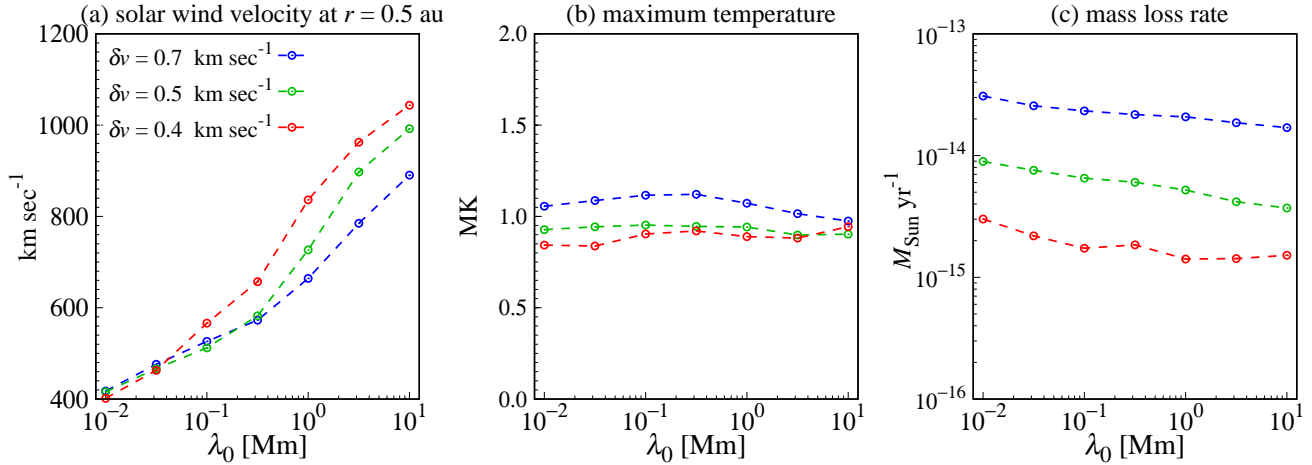} 
	\end{flushright}
	\vspace{1em}
	\caption{
		Time-averaged wind velocity at $r=0.5 {\rm \ au}$ (left), maximum temperature (middle), and mass loss rate (right) 
		for various $\lambda_0$ and $\delta v_0$.
		The units for the vertical axes are ${\rm km \ s^{-1}}$, ${\rm MK}$, and $M_{\rm Sun} \ \rm{ yr^{-1}}$, respectively, where $M_{\rm Sun}$ is the solar mass.
		The horizontal axis shows $\lambda_0$ in unit of ${\rm Mm}$.
		The red, green, and blue lines indicate $\delta v_0=0.4 {\rm \ km \ s^{-1}}$, 
		$0.5 {\rm \ km \ s^{-1}}$, and $0.7 {\rm \ km \ s^{-1}}$, respectively.
		}
	\vspace{2em}
	\label{fig:corr_dependence}
	\end{figure*}

\subsection{Dependences on $\lambda_0$ and $\delta v_0$}

	Following \citet{Cranm07}, we discuss the dependences of the asymptotic solar wind velocity, maximum coronal temperature and mass loss rate
	on $\lambda_0$ and $\delta v_0$.
	We show in Figure \ref{fig:corr_dependence} the time-averaged (a) wind velocity at $r=0.5{\rm \ au}$, (b) maximum temperature and (c) mass loss rate.
	
	According to \citet{Cranm07} and \citet{Verdi10}, the solar wind velocity tends to increase when $\lambda_0$ increases.
	This is observed in our result regardless of $\delta v_0$.
	\citet{Cranm07} and \citet{Suzuk06a} also argue that the wind velocity decreases as $\delta v_0$ increases.
	This is evident for large $\lambda_0$.
	As for small $\lambda_0$, the wind velocity is nearly constant around $\sim 400 {\rm \ km\ s^{-1}}$, 
	indicating that the solar wind is completely thermally driven \citep{Parke58,Hartl68}.
	The maximum temperature changes little with $\lambda_0$ and $\delta v_0$ as seen in \citet{Cranm07}.
	Especially, the maximum temperature is almost constant with respect to $\lambda_0$.
        This invariance is because the change in $\lambda_0$ is balanced by adjusting the fractional contributions from turbulence and shock heating.
        Indeed, the total heating rate per unit mass is nearly the same for different $\lambda_0$ values (Figure \ref{fig:heating_pergram_compare}).
        Suppose that $\lambda_0$ becomes smaller and turbulent heating is enhanced.
        Due to the increase in the coronal gas pressure, the plasma becomes more incompressible, leading to the reduction of compressible heating.
	The mass loss rate is also independent of $\lambda_0$, because it is almost independent of the heating location \citep{Hanst95}.
	Meanwhile, the mass loss rate sensitively depends on $\delta v_0$ \citep{Suzuk06b,Cranm07}; 
	the cases with $\delta v_0=0.7 {\rm \ km  \ s^{-1}}$ give approximately an order of magnitude larger mass loss rates than the cases with $\delta v_0=0.4 {\rm \ km  \ s^{-1}}$, 
	even though the energy input $\propto \delta v_0^2$ is only three times larger. 
	This is mainly because the reflection of the Alfv\'{e}n waves is suppressed for larger $\delta v_0$;  
	a larger $\delta v_0$ gives higher coronal density and hotter corona because of the larger heating.
	The denser corona suppresses the wave reflection in the chromosphere because the density difference between the photosphere and the corona is smaller; 
	the hotter corona suppresses the wave reflection in the corona because the scale height is larger there. 
	These effects make a larger fraction of the input energy transmitted to the corona and the solar wind region to drive denser wind \citep{Suzuk13}.
	To summarize, our simulations obtain the same trends as those obtained by \citet{Cranm07}.
        This is not trivial because we include compressibile waves in the corona and solar wind and use different values for $c_d$.

\section{Summary \& Discussion} \label{sec:summary_discussion}
	
	In this paper, we proposed a new 1D model of the coronal heating and solar wind acceleration.
	Our model is a hybrid model of a time-dependent compressible heating model \citep{Suzuk05} and an incompressible heating model \citep{Lione14}.
	Because both the compressible heating and incompressible heating are important in the corona and solar wind \citep{Matsu14}, 
	our model likely gives a better result compared to previous studies.
	As shown in Figure \ref{fig:result_compare}, the corona and the solar wind are formed regardless of the photospheric correlation length $\lambda_0$.
	The main heating mechanism is the turbulence heating (Figure \ref{fig:heating_pergram_compare}), especially when $\lambda_0 \lesssim 1 {\rm \ Mm}$.
	This does not necessarily mean that the compressibility is negligible.
	The fact that we obtain sufficient amounts of turbulence heating even for $\lambda_0 = 1{\rm \ Mm}$ despite small $c_d$ ($c_d=0.1$) indicates that the compressibility plays a role in the enhancement of 
	Alfv\'en wave reflection and, as a result, turbulence heating via the parametric decay instability.
	The density fluctuation, cross helicity, and solar wind velocity are strongly affected by $\lambda_0$ (Figure \ref{fig:result_compare}, \ref{fig:ch_dro_compare}), 
	while the maximum coronal temperature and mass loss rate are almost independent of $\lambda_0$ (Figure \ref{fig:corr_dependence}).
		
	Our turbulence model is one of the simplest models currently in use, and there is much space for improvement.
	The turbulence term we used is derived from the reduced MHD equations \citep{Hossa95,Dmitr02}.
	However, this treatment may not be applicable to our compressible MHD system.
	For a better modeling, the direct treatment of the compressible MHD turbulence \citep{Grapp93,Chand05} may be a better approach.
	To discuss the spectral evolution in the wave-number space, a shell model \citep{Buchl07,Verdi12b} is a possible extension to solve a large range of spatial scales. 
		
	The photospheric correlation length, $\lambda_0$, should in principle be determined from the physical properties of the magneto-convection below the photosphere. 
	According to \citet{Rempe14}, the spatial power peak of the photospheric kinetic energy lies around $\lambda \sim 1{\rm \ Mm}$, 
	while the magnetic energy is nearly flat for $0.1 {\rm \ Mm} \lesssim \lambda \lesssim 10 {\rm \ Mm}$.
	$\lambda \sim 1{\rm \ Mm}$ is on the scale of granulation while $\lambda \sim 0.1 {\rm \ Mm}$ corresponds to the scale of intergranular sub-arcsec magnetic patches \citep{Berge95,Berge01}.
	If we simply assume that $\lambda_0$ is comparable to the energetic scale of the photospheric motion, $\lambda_0=1{\rm \ Mm}$ is a reasonable choice.
	This assumption is equivalent to the idea that the swaying motion of the flux tube is the main driver of transverse waves \citep{Stein98}.
	However, if the vortex flow inside the magnetic patches is the most energetic driver of transverse waves \citep{Balle11},  $\lambda_0=0.1{\rm \ Mm}$ is preferable.
	Radiation MHD simulations indicate that the vortex size in the chromosphere is around $\lambda \sim 1{\rm \ Mm}$ \citep{Moll012,Iijim17}.
	Therefore, the most preferable value of $\lambda_0$ lies between $0.1 {\rm \ Mm}$  and $1 {\rm \ Mm}$, depending on the generation mechanism of waves.
	As discussed in Section \ref{subsec:qss}, the coronal observation is consistent with $\lambda_0 \gtrsim 1 {\rm \ Mm}$.
	Thus, $\lambda_0 \sim 1 {\rm \ Mm}$ is consistent with the photospheric wave driving and coronal wave observation, 
	although the best choice for $\lambda_0$ should be investigated in detail from both aspects in future works.
		
	M.S. is supported by Leading Graduate Course for Frontiers of Mathematical Sciences and Physics (FMSP) and  Grant-in-Aid for JSPS Fellows.
	T.Y. is supported by JSPS KAKENHI Grant Number 15H03640.
	T.K.S. is supported in part by Grants-in-Aid for Scientific Research from the MEXT of Japan, 17H01105.
	Numerical calculations were in part carried out on PC cluster at Center for Computational Astrophysics, National Astronomical Observatory of Japan.

\end{document}